# Ion transport study of mechanically-milled amorphous AgI-Ag$_2$O-V$_2$O$_5$ fast ionic conductors

Parveen Kumar[†] & K Shahi[†]*

[†]Solid State Ionics Laboratory, Advanced Center for Materials Science, Indian Institute of Technology Kanpur, India -208016
*Department of Physics Indian Institute of Technology Kanpur, India -208016
Email: parveen.kumar.2006@alumni.iitk.ac.in

The structural and electrical characterizations of mechanically-milled (MM) amorphous fast ionic conductors (a-FICs), viz. $x$AgI (100-$x$)[0.67 Ag$_2$O-0.33V$_2$O$_5$] ( $x$ = 40, 50, 55 and 70 ) have been reported. The amorphisation is restricted only to the compositions which are well within the glass forming region and all samples are found to be highly agglomerated and $X$-ray amorphous in nature. The frequency dependent $ac$ conductivity, $\sigma'(\omega)$, of the amorphous samples investigated in the frequency range 5Hz -13 MHz and temperature range 100- 350 K shows a $dc$ conductivity regime at low frequencies and a dispersive regime at higher frequencies. The $\sigma'(\omega)$ spectra can be described by the Jonscher power law (JPL), $\sigma'(\omega) = \sigma_{dc} + A(T)\,\omega^n$. However, the values σ$_{dc}$($T$) and A($T$) both show two distinct Arrhenius regions and $n$ ( < 1 ) is found to be temperature dependent, i.e., decreasing with increasing temperature.

## 1. Introduction

In the last few years the mechanochemical synthesis has been found to be an effective technique[1-4] for the synthesis of amorphous fast ionic conductors (a-FICs) in the field of solid state ionics. The mechanically-milled (MM) glasses are found to be almost similar to the melt-quenched (MQ) glasses. For example, both MM and MQ glasses show similar $X$-ray diffraction pattern[2] and high value of conductivity, a glass transition temperature and crystallization temperature ($T_c$). Further, the electrochemical properties of MM and MQ glasses are also similar[2]. However, there are subtle but significant differences between MM and MQ type of disordered solids, such as MM glasses show two crystallization temperatures instead of one as shown by MQ glasses. Moreover, MM glasses show two regimes of activation in their σ$_{dc}$ versus 1/$T$ plot, a low temperature region with lower activation energy and a high temperature region with higher activation energy[3]. In view of these differences between the two types of disordered solids and due to importance of ion dynamics study[4-5], the MM glasses are viewed as an important class of materials for further investigation.

Recently, the $ac$ conductivity studies on mechanically-milled Ag$_2$S-based amorphous fast ion conducting glasses[3] have been reported. The $ac$ conductivity of these new types of disordered solids is found to be similar to that of the MQ glassy fast ionic conductors, which is described by the well known Jonscher power law (JPL);

$$\sigma'(\omega) = \sigma_{dc}(T) + A(T)\,\omega^n \qquad \ldots (1)$$

where the $dc$ conductivity ( σ$_{dc}$ ) and the parameter $A$ are frequency-independent and obey Arrhenius temperature dependence and are given, respectively, by

$$\sigma_{dc}(T) = \sigma_0 \exp\left(-\frac{E_{dc}}{kT}\right) \qquad \ldots (2)$$

$$A(T) = A_0 \exp\left(-\frac{E_{ac}}{kT}\right) \qquad \ldots (3)$$

where σ$_0$ and $A_0$ are constants, and $E_{dc}$ and $E_{ac}$ are $dc$ and $ac$ activation energies, respectively and frequency exponent $n$ in Eq. (1) is supposed to be a material property and the following relation is satisfied

$$E_{ac} = (1-n)\,E_{dc} \qquad \ldots (4)$$

However, $A(T)$ shows two Arrhenius regions instead of one as observed in MQ glasses and even more, the frequency exponent $n$ is a function of temperature, i.e., decreasing with increasing temperature. The relation $E_{ac} = (1-n)\,E_{dc}$ is satisfied only at lower temperatures[3].

For further investigation of the differences between MM and MQ glasses and to establish the unique properties of MM glasses, we now report the $ac$ conductivity studies on another mechanochemically synthesized amorphous fast ionic conductor, i.e., AgI-Ag$_2$O-V$_2$O$_5$.

## 2. Experimental Details

The nominal compositions such as 40 AgI. 40 $Ag_2O$. 20 $V_2O_5$, 50 AgI. 33 $Ag_2O$. 17 $V_2O_5$, 55 AgI. 30 $Ag_2O$. 15 $V_2O_5$, 70 AgI. 20 $Ag_2O$. 10 $V_2O_5$ abbreviated as 40 SISOVO, 50 SISOVO, 55 SISOVO and 70 SISOVO, respectively are chosen for the synthesis and characterization. The first two compositions lie well within the glass forming region as per the phase diagram[6], whereas the third one is on the boundary and the fourth lies out side the glass forming region. The samples of desired compositions are ball milled in agate pot of 250 ml volume with 4-6 ball of 20 mm diameter in acetone medium. The acetone level is maintained constant during the ball-milling process. The samples are milled for 18-90 hr depending on the AgI content in the sample, to produce the amorphous phase. After milling, the acetone is removed through vacuum drying and fine powders thus obtained, are directly used for further characterization. The XRD studies are done through Rich.Seifert MZ III and SEM studies through QUANTA-200. The electrical characterization is done on the palletized samples[3] with the help of HP 4192 A impedance analyzer in a cryostat of Lab-Equip (India) in the temperature range 100–350 K and frequency range 5 Hz -13 MHz.

## 3. Results and Discussion

The XRD results of the mechanically-milled SISOVO samples are very similar to those reported earlier[2]. Fig.1 shows a typical XRD pattern for 55 SISOVO MM sample milled for different durations as well as for MQ sample along with that of the unprocessed (0 hr) material. It is evident from the XRD pattern that the intensity of the peaks corresponding to different constituents of the unprocessed sample decreases with increasing milling time and eventually disappears after complete amorphisation (20-27 hr for 55 SISOVO). Further, there is hardly any difference between the XRD patterns of MM and MQ glasses. However, the 70 SISOVO sample which lies out side the glass forming region does not amorphize completely even after a prolonged (90 hr) ball-milling.

Figs 2(a and b) show the high resolution electron micrographs of the unprocessed and processed 55 SISOVO samples. An agglomeration of particles is observed in the MM samples indicating that the actual size of the particles may be even smaller than that observed in these micrographs, and the processed and unprocessed samples have clear morphological differences.

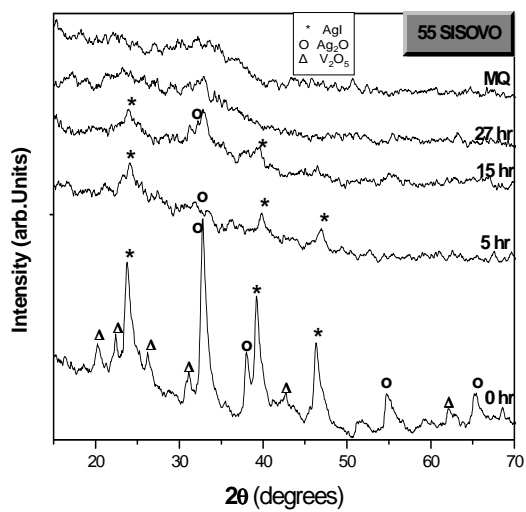

Fig. 1 *X*-ray diffraction (XRD) patterns of 55 SISOVO sample mechanically milled for different durations at room temperature. XRD pattern of melt quenched glass of the same composition is also shown. Symbols: (*) AgI, (o) $Ag_2O$, (Δ) $V_2O_5$ peaks

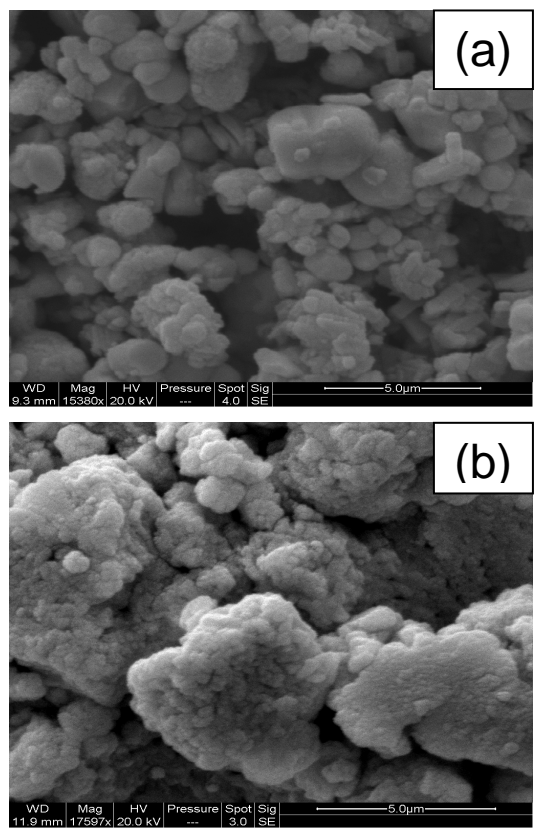

Fig. 2 High resolution scanning electron micrographs for (a) unprocessed 55 SISOVO and (b) 27 hr MM 55 SISOVO sample

Fig. 3 shows a typical *ac* conductivity spetctra for 55 SOSOVO. It is similar to that of the other a-FICs reported earlier by us[3] and other researchers working on ion dynamics[4-5, 8]. The *ac* conductivity spectra for 55 SISOVO comprise a low frequency *dc* conductivity region and a high frequency dispersive region. The Jonscher power law [JPL, Eq.(1)] reproduces the *ac* conductivity spectra well for 55 SISOVO. The solid lines are the best-fit curves according to JPL (Eq. 1). The best-fit values of the parameters $\sigma_{dc}$, $A$ and $n$ to the *ac* conductivity spectra (Fig. 3) for different compositions (40, 50, 55 and 70 SISOVO) are shown in Figs 4-6.

The best-fit values of $\sigma_{dc}$ obtained from the fitting of Eq.(1) to the *ac* conductivity spectra of SISOVO vary Arrheniously with temperature as shown in Fig 4. However, unlike melt quenched glasses[5], the *dc* conductivity of mechanically-milled SISOVO samples shows two Arrhenius regions as observed earlier for the $Ag_2S$-based system[3]. The $\sigma_{dc}$ versus $1/T$ shows a low temperature linear region with lower activation energy and a high temperature region with slightly higher activation energy (Table 1) and the difference between two *dc* activation energies decreases with increasing AgI content. The knee temperature separating the two Arrhenius regions decreases with the increasing concentration of the dopant salt, i.e. AgI.

Fig.5 shows an Arrhenius representation of $A$, a fitting parameter of Eq.(1), to the *ac* conductivity spectra of various compositions of SISOVO system. The $A$ versus $1/T$ plots also show two Arrhenius regions[3, 7] separated by a knee temperature ($T_N$) which shifts towards lower temperature with increasing content of the dopant, i.e., AgI. The knee temperature ($T_N$) separating the two regions in $\sigma_{dc}(T)$ and $A(T)$ are not very close to each other which is in agreement with earlier observations[3]. These results suggest that both have the same origin[7]. In view of the two Arrhenis regions in $\sigma_{dc}(T)$ and $A(T)$ along with the presence two crystallization temperatures[2] in DSC studies, it may be suggested that these new type of disordered a-FICs are also associated with two sets of $Ag^+$ ions[3]. The slope of $A$ versus $1/T$ yields the *ac* activation energy for hopping of ions.

The variation of fitting parameter $n$ of Eq. (1) to the conductivity spectra for various compositions of SISOVO, with temperature is shown in Fig. 6. The value of exponent decreases slowly initially but more

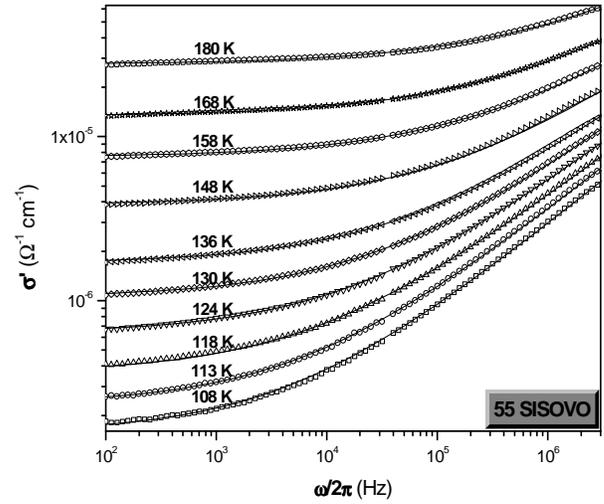

Fig. 3 Frequency dependence of conductivity at different temperatures (108-180 K) for 55 SISOVO sample

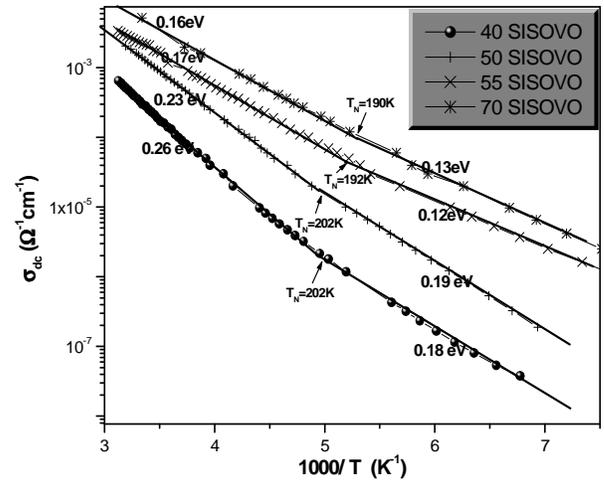

Fig. 4 Variation of $\sigma_{dc}$, calculated from the conductivity isotherms $\sigma'(\omega)$ as a function of $1/T$ for 40, 50, 55 and 70 SISOVO samples

Table 1 *dc* activation energy ($E_{dc}$) obtained from log ($\sigma_{dc}$) versus $1/T$ and *ac* activation energy ($E_{ac}$) obtained from the log $A$ versus $1/T$ plot in the low and high temperature regions for 40, 50, 55 and 70 SISOVO samples

| System | $E_{dc}$ (eV) | | $E_{ac}$ (eV) | |
| | $\sigma_{dc}$ versus $1/T$ | | log $A$ versus $1/T$ | |
| | 125-200K | 200-330K | 125-180K | 180-230K |
| 40 SISOVO | 0.18 | 0.26 | 0.11 | 0.35 |
| 50 SISOVO | 0.19 | 0.23 | 0.10 | 0.22 |
| 55 SISOVO | 0.12 | 0.17 | 0.08 | 0.23 |
| 70 SISOVO | 0.13 | 0.16 | 0.09 | 0.21 |

Table 2 *dc* activation energy ($E_{dc}$), *ac* activation energy ($E_{ac}$) calculated from Eq. (4) and observed in the plot of log *A* versus $1/T$ in the temperature range 125-180K for 40, 50, 55 and 70 SISOVO samples

| System | $E_{dc}$ (eV) | Average value of *n* | $E_{ac}$ (eV) Calculated | $E_{ac}$ (eV) Observed |
|---|---|---|---|---|
| 40 SISOVO | 0.18 | 0.53 | 0.09 | 0.11 |
| 50 SISOVO | 0.19 | 0.47 | 0.10 | 0.10 |
| 55 SISOVO | 0.12 | 0.33 | 0.08 | 0.08 |
| 70 SISOVO | 0.13 | 0.30 | 0.09 | 0.09 |

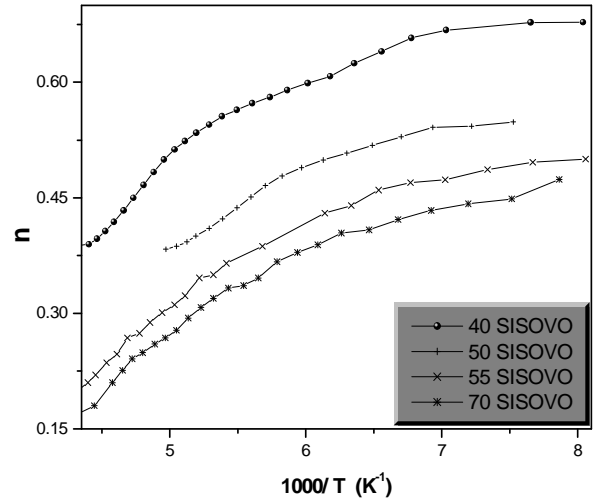

Fig. 6 The variation of frequency exponent factor *n* in the JPL Eq. (1) as a function of $1/T$ for 40, 50, 55 and 70 SISOVO samples.

## Conclusions

These results are analogous to our previous observations on the some other systems[3]. The observations indicate that the mechanochemical synthesis can produce *X*-ray amorphous phases within the glass forming window. The SEM analysis indicates the presence of highly agglomerated particles which also reveal that the presence of nanosize particles cannot be ruled out. The *ac* conductivity spectra obeys the JPL, however, the presence of two Arrhenius regions in the $\sigma_{dc}(T)$ and $A(T)$ along with the two crystallization temperatures[2] in these a-FICs suggest the presence of two types of bonding states of the $Ag^+$ ion in the amorphous matrix.

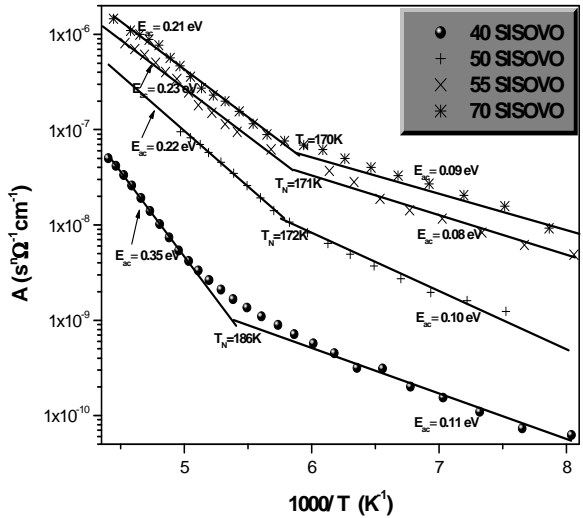

Fig. 5 Variation of parameter *A* in JPL Eq. (1) as a function of $1/T$ for 40, 50, 55 and 70 SISOVO samples

rapidly at higher temperatures for all the compositions of SISOVO. The rapid decrease of *n* at higher temperatures may be due to window effect[8] or due to less dispersive contribution present in the conductivity spectra at higher temperatures.

Moreover, the relation between $E_{ac}$, $E_{dc}$ and *n* (Eq.4), as predicted by JPL, is fairly satisfied at lower temperatures (Table 2). However, this correlation does not hold good at higher temperatures.